

\def\AbstractBegins
{
 \singlespace                                        
 \bigskip\leftskip=1.5truecm\rightskip=1.5truecm     
 \centerline{\bf Abstract}
 \smallskip
 \noindent
 } 
\def\AbstractEnds{\bigskip\leftskip=0truecm\rightskip=0truecm}

\def\ReferencesBegin
{
\singlespace					   
\vskip 0.5truein
\centerline           {\bf References}
\par\nobreak
\medskip
\noindent
\parindent=0pt
\parskip=4pt			        
 }

\def\section #1    {\bigskip\noindent{\bf  #1 }\par\nobreak\smallskip}

\def\subsection #1 {\medskip\noindent{\it [ #1 ]}\par\nobreak\smallskip}



%
 \let\miguu=\footnote
 \def\footnote#1#2{{$\,$\parindent=9pt\baselineskip=13pt%
 \miguu{#1}{#2\vskip -5truept}}}
%
\def\Integers{\hbox{${\rm Z \kern 0.3ex \llap{Z}}$}}
 

\def\=>{\Rightarrow}
\def\==>{\Longrightarrow}
 
 \def\dal{\displaystyle{{\hbox to 0pt{$\sqcup$\hss}}\sqcap}}
 
%
\def\lto{\mathop
        {\hbox{${\lower3.8pt\hbox{$<$}}\atop{\raise0.2pt\hbox{$\sim$}}$}}}
\def\gto{\mathop
        {\hbox{${\lower3.8pt\hbox{$>$}}\atop{\raise0.2pt\hbox{$\sim$}}$}}}
%
 


\def\tr{\rm tr}			


\def\to{\rightarrow}		

\def\hat{\widehat}		


\def\ideq{\equiv}		


\def				
  \Complexes
   {{\rm C}\llap{\vrule height6.3pt width1pt depth-.4pt\phantom t}}




\def\interior #1 {  \buildrel\circ\over  #1}     




\message{--> VERSION 1.2 of the manuscript}



\message{-- Assuming 8.5 X 11 inch paper}

\magnification=\magstep1		
\raggedbottom

\hsize=6.4 true in
 \hoffset=0.27 true in		
\vsize=8.7 true in
 \voffset=0.75 true in		

\parskip=9pt

\def\singlespace{\baselineskip=12pt}  

\def\sesquispace{\baselineskip=15pt}                 


\def\Hext{{\cal H}_{ext}}
\def\lam{\lambda}



\rightline{gr-qc/9701056}

\vfill
\bigskip

\centerline{ {\bf How Wrinkled is the Surface of a Black Hole?}\footnote{*}%
 {To appear in David Wiltshire (ed.), Proceedings of the First Australasian
 Conference on General Relativity and Gravitation, held February, 1996,
 Adelaide, Australia}}



\singlespace			        


\bigskip
\centerline {\it Rafael D. Sorkin}
\medskip


\smallskip
\centerline {\it Instituto de Ciencias Nucleares, 
                 UNAM, A. Postal 70-543,
                 D.F. 04510, Mexico}


\smallskip
\centerline { \it and }
\smallskip
\centerline {\it Department of Physics, 
                 Syracuse University, 
                 Syracuse, NY 13244-1130, U.S.A.}
\smallskip
\centerline {\it \qquad\qquad internet address: sorkin@nuclecu.unam.mx}

\AbstractBegins 
We present evidence that, below a certain threshold scale, the horizon of a
black hole is strongly wrinkled, with its shape manifesting a self-similar
(``fractal'') spectrum of fluctuations on all scales below the threshold.
This threshold scale is small compared to the radius of the black hole, but
still much larger than the Planck scale.  If present, such fluctuations
might account for a large part of the horizon entropy.
\AbstractEnds

\sesquispace
\bigskip\medskip

\section{ Introduction }

Can a calculation based on Newtonian gravity teach us anything about a
black hole?  If it can, then we will see that the surface of a black hole
must be strongly wrinkled on scales below a certain threshold scale
$\lambda_0$, which in a certain Newtonian approximation comes out as
$(M{}l_p^2)^{1/3}$, $l_p$ being the Planck length.  It also looks plausible
that this wrinkling would be self-similar, lending the horizon what might
be called a ``fractal'' shape.

Such a departure from smoothness of the event horizon, seems noteworthy in
itself, but probably its greatest significance would be in connection with
black hole thermodynamics.  Let us therefore take a few moments to review
some of the open questions in that subject.
One knows from a preponderance of evidence that a black hole behaves as if
its horizon carried a ``surface entropy'' of $2\pi{}A/l_p^2$ (where
$l_p^2=8\pi{}G\hbar$, in units with $c=1$).  Most of this evidence pertains
to stationary black holes (the ``First Law'' relating variations in a black
hole's mass to variations in its horizon area, the instanton computations
of the partition function at ``tree level'', the thermal radiance), but
there is also the ``Second Law'' or area increase theorem (proved assuming
``cosmic censorship''), which applies to black holes out of equilibrium.
This result can be interpreted as the $\hbar\to{}0$ limit of the
thermodynamic Second Law for systems including black holes (In that limit
the black hole hole entropy should overwhelm all other contributions, since
the Planck length goes to $0$ while $S\sim{}A/l_p^2$.), and on this
interpretation, the horizon area gives the entropy even for black holes
which are far from stationary.

(It is sometimes suggested that one should identify the surface of a black
hole with its {\it apparent horizon}, and that therefore the entropy of a
black hole away from equilibrium ought to be the area of its apparent
horizon, rather than that of its event horizon.  There is even a form of
area theorem for the apparent horizon [1].  However, an
entropy based on the apparent horizon would sometimes jump discontinuously,
although no physical discontinuity in the metric or other fields would be
present.  Moreover the concept of event horizon is more robust than that of
apparent horizon, and can make sense even where the notion of smooth curve,
or the divergence of a vector field does not.  In particular the notion of
event horizon still makes sense in the context of causal sets, since it
relies only on the existence of a causal order.)

Despite all the evidence for the existence of an entropy associated with
black hole horizons, and despite the evidence that the resulting total
entropy (horizon area plus exterior entropy) is non-decreasing, there is
still very little understanding of the ``statistical mechanical
explanation'' of these facts [2].  In particular a derivation
of entropy increase on the usual pattern would have to rely on features,
like unitarity, ergodicity and weak coupling, which appear to be absent in
the case of black holes; moreover, such a derivation would have to overcome
the serious objection that, even for a black hole near equilibrium, its
{\it interior} region is far from stationary, rendering very doubtful the
assumption that the number of internal states contributing to the black
hole's entropy can be deduced just from a few external parameters such as
its mass and angular momentum.  (For more discussion of these points see
[3] [4].)  For these reasons, among others, it
remains unclear what degrees of freedom the black hole entropy refers to:
what states are being counted by $N$ when one writes $S=\log N$ for a black
hole.

\section{ How the Second Law might be proved }

  If one reflects that the (``Generalized'') Second Law refers effectively
only to the region external to the black hole, and if one takes to heart
the fact that this region ought (by the very definition of a black hole!)
to obey an essentially ``autonomous'' dynamics of its own, then it becomes
natural to seek a proof of entropy increase based on the ``coarse-graining''
that consists in ignoring whatever is occurring beyond the veil of the
event horizon.

  I have earlier proposed such a proof [3], or rather a
proof-scheme which can be filled in within any theory of quantum gravity
that incorporates certain basic features.  These features are:

\item{$\bullet$}        that the (mixed) state for the external region be
describable by an effective density operator $\rho_{ext}$ acting in some
hilbert space $\Hext$

\item{$\bullet$}         that $\rho_{ext}$ obey a  law of evolution
which is (at least to a good approximation)  autonomous and ``Markovian''

\item{$\bullet$}   that energy be conserved and given by an operator
$E$ defined in the external region and acting in $\Hext$

\item{$\bullet$}        that the subspace of $\Hext$
in which $E<E_0$ be {\it finite-dimensional} for any fixed energy 
bound $E_0$. 

\noindent
It then follows rigorously that the value of $S:=\tr\rho\log{}\rho^{-1}$
cannot decrease as the hypersurface to which it refers moves forward in
time.  (For maximum generality, this hypersurface should be defined in some
a priori manner (allowing reference to the rigid ``box'' in which the whole
system is taken to be enclosed).)

Notice that this approach to proving the Second Law requires that the
degrees of freedom which the black hole entropy manifests be accessible in
the region {\it outside} of the black hole (modulo whatever small blurring
of the horizon can be expected from quantum gravitational effects).

\section{ Possible sources for the entropy (and a gas analogy) }

Aside from locating them in the external region, the proof just discussed
does not specify what ``degrees of freedom'' the information represented by
the horizon entropy actually expresses.  Ideally one would like to trace
these degrees of freedom directly to some fundamental quantum theory
underlying General Relativity.  Thus one might seek them in the causal
links straddling the horizon (causal set theory [5]), in the
``fundamental loops'' straddling the horizon (loop representation in
canonical quantum gravity), or in the variables describing the fundamental
strings intersecting the horizon (superstring theory\footnote{*}%
{Note added later: I do not know whether the various $branes$ recently
proposed as carriers of the entropy in superstring theory can be regarded
as localized in the neighborhood of the horizon or not.}).

No doubt, understanding the entropy in this way would teach us most about the
nature of quantum gravity on microscopic scales.  It should in particular
answer the question whether spacetime exists at all fundamentally, and if
not, what replaces it.  However, it is also conceivable that the
fundamental variables, whatever they might be, admit of an effective
description on super-Planckian scales, in terms of which much or all of the
entropy could be described in terms of currently understood theory.  Such a
description might not teach us as much about the deeper nature of
spacetime, but it would be equally valid in its proper domain.  And it
would represent for the deeper theory a signpost, and a challenge to
connect up the deeper degrees of freedom with the more phenomenological ones.

For the sake of analogy, consider a box of gas at high temperature.  Here,
the fundamental degrees of freedom are those of the molecules composing the
gas: their positions and (in the case of a classical description) their
velocities.  Fundamentally the finiteness of the entropy of such a gas
rests on the finiteness of the number of particles composing it, as can be
seen if the entropy is written as follows:
$$
       S/k = N \log {V T^{d/2} \over M K_0}.   \eqno(1)
$$
Here $d$ is the (spatial) dimension, $N$ the number of particles, $V$ the
volume, $T$ the temperature, $M$ the total mass of the gas, and $K_0$ is a
constant depending on Boltzmann's constant $k$, $\hbar$, and the mass of an
individual molecule, specifically $K_0$ is a $d$-dependent numerical factor
times $\sqrt{\hbar^{2d} / k^d m^{d+2}}$.  The formula is an approximation
which is valid as long as the gas is hot enough to avoid quantum degeneracy,
i.e. as long as the argument of the logarithm is $\gg{}1$.

It is clear from this formula that the entropy goes to infinity with $N$.
This does not mean, however, that it would be impossible to understand (at
least part of) the entropy in a continuum picture.  Indeed fluctuations in
the molecular positions can be redescribed --- if they are sufficiently
regular --- as fluctuations in a continuum density function $f$, and more
generally, the gas can be described at some level of approximation as a
fluid.  If an entropy could be computed within a fluid description, then
one would obtain an accounting of the contributing micro-states in terms of
the fluctuations of collective degrees of freedom like the mass-density and
velocity fields.  Computed this way, the entropy would presumably come out
infinite for a truly continuous fluid (which wouldn't know about the size
of a molecule), but it could be rendered finite by omitting the physically
meaningless density fluctuations occurring (in the continuum model) below
the scale set by the intermolecular spacing.  The question would be then:
how much of the entropy would we recover in this manner?

I don't know if anyone has done such a computation, but I would like to
take this opportunity to comment on some aspects of the problem, limiting
myself to the case of a non-degenerate, dilute gas.  If one thinks to
quantize the ``sound waves'' of the gas (the irrotational modes of the
linearized fluid equations), and if one ignores the damping of these waves,
one obtains, naturally, a typical black body --- or in this case ``silent
body'' --- spectrum of phonons, with associated finite entropy.  Without a
cutoff, this entropy is much in excess of the correct answer
(1), but if one cuts the sound modes off at the intermolecular
spacing, then the entropy comes out nearly correct(!).

But really, there are at least three relevant length-scales in this
problem: the molecular mean-free-path, the intermolecular spacing, and the
de Broglie wavelength of a molecule, each much bigger than the next (for a
non-degenerate, dilute gas).  Logically, one should take the first rather
than the second of these as the sound cutoff, because below that scale
phonons clearly cannot propagate.\footnote{*}
{The third length-scale, namely the molecular de Broglie wavelength, plays
a role formally as the shortest wavelength of phonons which can be
thermally excited at the given temperature.  That is, it provides the
familiar quantum cutoff that renders the ``silent body'' entropy finite,
despite the infinity of phonon modes that formally exist, in the absence of
damping.  When the gas is just verging on quantum degeneracy, this third
length-scale approaches the intermolecular spacing, and the silent-body
entropy takes on the correct order of magnitude without the need of a
cutoff.}
This shows up in the continuum approximation as a wavelength-dependent
damping of the sound modes which becomes a critical damping when the
wavelength reaches the mean free path.  Thus, the ``silent body phonon
entropy'' is actually much {\it less} than the full entropy, when the
non-propagating modes are omitted.  

So, how can we estimate the contribution of these omitted modes?  In the
continuum approximation without any cutoff, irrotational modes exist with
wavelengths right down to zero, but those shorter than the mean free path
have purely imaginary frequencies (they are non-propagating).  Therefore,
in order to evaluate the omitted modes' contribution to the entropy, we
would have to understand the entropy of a {\it damped} harmonic oscillator.
It seems plausible that such an oscillator carries more entropy than an
undamped one, and in particular that it has entropy, even in its ground
state.  If so, the entropy from the {\it propagating} sound modes would 
increase by an amount to be determined, but more importantly, one might
hope that the non-propagating modes lying between the mean free path and
the intermolecular spacing would still contribute the entropy required to
produce the correct total.  On the other hand, the infinity of
non-propagating modes existing {\it below} the intermolecular spacing might
by the same token be expected to contribute an infinite entropy, confirming
the expectation that a finite total entropy demands a finite cutoff.
(Might a similar infinite entropy be produced by the infinity of highly
damped, quasi-normal modes of a black hole?)

In addition to the irrotational modes, there are (except in $d=1$) a huge
number of rotational ones, which the above discussion has totally
neglected.  Such ``vortex modes'' offer another source of entropy beyond
that of sound, a source whose contribution is also plausibly infinite in
the continuum theory (especially since there seems to be no reason for the
frequencies of such modes to grow with decreasing wavelength).  And further
complicating matters are the nonlinear terms in the fluid equations, whose
presence might invalidate any computation of the entropy carried out within
a purely linear approximation.

A final comment here is that the phenomenological parameters (or ``coupling
constants'') which enter the fluid equations, such as the viscosity, the
heat conductivity and the speed of sound, implicitly contain information
about the values of microscopic quantities such as the molecular mass.
Thus, my earlier argument that the fluid model ``wouldn't know about the
size of a molecule'' was at best suggestive; and only a more careful
analysis of the sort just sketched can tell us for sure whether a cutoff is
required to render the entropy finite.

Now let us contemplate a black hole in the spirit of the above discussion.
Is it possible that all or part of its entropy can be accounted for in
terms of effective degrees of freedom which are independent of whatever
variables a deeper theory might prescribe, for example the degrees of
freedom of the standard model, including gravity?  Here I wish to consider
only two possible contributions, both of which will turn out to be
intimately related with the horizon wrinkling toward which we are heading.

\subsection {Entropy as shapes}

The first of these possibilities is perhaps the most obvious one
[6], namely that the $e^S$ microscopic alternatives the
entropy is counting are the alternative {\it shapes} of the horizon.  This
explanation is appealing because it offers a geometrical origin for the
very geometrical relationship
$$
                    S = 2\pi A.           \eqno(2)
$$
(In fact, even the factor of $2\pi$ in this equation is geometrical!  It
represents the radius of the unit circle in one way of doing the tree level
instanton calculation [7].)  The universality of the
coefficient in this equation would thereby be traced to the universality of
the geometrical degrees of freedom of the horizon, which are always the
same, independently of whatever non-gravitational fields may be present in
the theory.

\subsection {Entropy as entanglement}

A second possibility [8] (not necessarily exclusive to the
first) is that the entropy is carried by quantum fields propagating in the
neighborhood of the horizon, or more specifically that it is the ``entropy
of entanglement'' which arises when one neglects the correlations between
the field just inside and just outside the horizon, i.e. when one performs
the coarse-graining referred to earlier in connection with the Second Law.
This entanglement entropy can be computed for the case of a free field
[6] [9], and, as suggested by our gas analogy, it
turns out to be infinite in the absence of a cutoff.

Without going into the details, one can still give an intuitive picture of
the origin of this infinity.  Let us imagine a fluctuation in the field
$\phi$ of linear extension $\lambda$ in the neighborhood of the horizon.
If the fluctuation is totally outside or totally inside the horizon, it
contributes no more to the entropy than it would to the entropy of the
vacuum in flat spacetime.  But if it happens to sit astride the horizon,
then it sets up a correlation in the value of $\phi$ between inside and
outside, which is the ``entanglement'' that gets lost when one traces out
the degrees of freedom inside the horizon.  Since field fluctuations can
occur independently on arbitrarily small scales, one can understand that
their total contribution to $S$ is infinite.

When a cutoff is imposed, one gets instead of infinity, the result 
$S=(cst.)A/l_0^2$, where $l_0$ is the cutoff expressed as a length, and the
constant is of order unity, its precise value depending on how the cutoff
is introduced and normalized.  ($S$ can also have corrections of higher
order in the ratio of the cutoff to the radius of curvature of the
horizon.)  But this value for $S$ has the right order of magnitude
precisely when $l_0\sim{l_p}$.  Given this striking result, it is tempting
to conclude first, that the horizon entropy is indeed entanglement entropy,
and second, that its finiteness is telling us about a fundamental
granularity of spacetime.

\subsection {Species dependence and the coupling of field fluctuations to 
            the horizon } 

There is, however, at least one worry which at first sight would seem to
prevent the identification of horizon entropy with the entanglement entropy
of fields, namely the so-called ``species-dependence problem'', that is,
the problem that the precise magnitude of the entanglement entropy would
seem to depend on the number and type of fields present in nature, whereas
the formula (2) cares only about the area of the horizon.

One might think that this difficulty could be avoided only thanks to some
inbuilt constraint on which fields actually exist in nature (as might occur
in a unified theory such as superstring theory), or alternatively that it
could be avoided by ``back-reaction'' effects which would couple the field
fluctuations to the horizon shape, thereby modifying the formula for the
entanglement entropy.  It is actually the second idea which motivated the
calculation I want to describe in a moment; but, interestingly enough, we
now know that there might not be any difficulty at all, thanks to the work
of [10], which pointed out that the (renormalized) value of $G$
{\it also} depends on the number of species, and in just the way needed to
cancel the species dependence of the entropy.  It is true that adding (say)
a new species of particle will necessarily increase the entropy {\it at
fixed cutoff}.  But, at fixed cutoff the value of $1/8\pi{G}$ {\it also}
will be modified by the addition of a new species, i.e. $l_p$ will be
modified; and an effective-Action calculation then indicates that the two
modifications will compensate each other, so that the relation
$S=2\pi{A}/l_p^2$ will remain unaffected.

Although this is heartening, it does not return the back-reaction genie to
her bottle: we still have to sort out how field fluctuations distort the
horizon's shape, if we want to understand the status of the entanglement
entropy.  It might turn out that field fluctuations coupled strongly to the
horizon shape for wavelengths $\lambda$ below some $\lambda_0$, and if it
did, then the attendant deformations of the horizon could not be ignored
(as they have been so far) in computing the entanglement entropy.  How then
can we estimate the strength of the coupling between the horizon and the
quantum field fluctuations in its neighborhood?

\section { A Newtonian calculation of the induced fluctuations in 
           the horizon's shape }

To get at least a preliminary indication of when this coupling is likely to
be important, let us estimate it [11] in the crudest possible
manner, namely using a Newtonian approximation for the gravitational field
produced by the field fluctuations.  Since Newtonian gravity is so easy, I
can give the calculation in full (taking $c\ideq{1}$ and $8\pi{G}\ideq{1}$).

To get started, we need a definition of the Newtonian horizon, and I will
use the usual one, which locates it at the surface where the escape
velocity is that of light, i.e. at the locus of points where $v=1$ in the
equality $ mv^2/2+mV=0$, $V$ being the Newtonian potential.  (Notice,
however that this definition totally ignores any time-dependence in the
gravitational field.)  The equation defining the horizon is then
$$
           V = -1/2 .
$$
For our unperturbed horizon, we take that of a point mass $M$, which turns
out to be the sphere whose radius $R$ is (by a famous coincidence) exactly
the Schwarzschild radius, $R=M/4\pi$.  Thus, our unperturbed gravitational
potential, when expressed in terms of $R$, is
$$
          V_0 = - { R \over 2r}
$$
where $r$ denotes the distance to the center of the black hole.  

Now consider a fluctuation of size $\lambda$ which happens to find itself
astride the horizon.  On dimensional grounds, its associated energy should
be of order $1/\lam$, so I will take it to be  $m=4{\pi}f/{\lam}$
where $f$ is a conveniently normalized ``fudge factor'' of order unity.
The energy $m$ will be spread out over the support of the fluctuation
somehow, but the precise density profile will not affect our conclusions.
For convenience, I will use a density of $\rho=2f/r_1(r_1+\lam)^3$, where
$r_1$ is the distance to the center of the fluctuation.  The potential
caused by such a mass distribution is
$$
    V_1 = { -f / 2\lam \over r_1 + \lam }.
$$
Hence, the location of the perturbed horizon is determined by 
$$
    - 2 V  = { R\over r} +  { f/ \lam \over \lam  + r_1} = 1 \eqno(3)
$$
where $V=V_0+V_1$ is the total perturbed potential.

\subsection { The height and shape of the deformation, neglecting retardation}

To get an idea of the height of the bulge (or depression) induced by the
field fluctuation consider equation (3) ``on axis'', i.e. along a
radial line joining the center of the unperturbed black hole to the center
of the fluctuation (which we take to lie precisely on the unperturbed
horizon).  With $h:=R-r$ the height of the bulge, we have from (3)
$$
    {R \over h + R} +  { f/\lam \over \lam + h } = 1 .
$$
This equation is easily solved exactly, but it is just as instructive to
solve it in the approximation $h,\lam{\ll}R$, where it becomes
$$
    {h\over\lam} ( {h\over\lam} + 1 )  
       \approx {fR\over\lam^3} \ideq ( {\lam_0\over\lam} )^3
$$
with
$$
            \lam_0 = (f R)^{1/3} .
$$

{}From this it is easy to see that $\lam\sim\lam_0$ is a critical length,
above which a fluctuation of size $\lam$ induces only a very small $h$ such
that
$$
      h/\lam \sim (\lam_0 / \lam)^3 \ll 1 ;
$$
in other words, the distortion of the horizon is much smaller than the
fluctuation itself, and in this sense the coupling between them is weak.
For $\lam\sim\lam_0$, on the other hand, we have $h/\lam\sim{1}$, and the
distortion is comparable in size to that of the fluctuation which raised it
(strong coupling).  Finally, for $\lam\ll\lam_0$ we nominally find a bulge
which is much greater than the fluctuation size, but here our
approximations are clearly breaking down: it is no longer reasonable to
treat the fluctuation in isolation from other fluctuations, nor is it
reasonable in this regime to have neglected retardation effects, given the
finite lifetime of the field fluctuation.

Finding the profile of the induced bulge (or depression) is also
straightforward.  With the same approximations as before, we can treat the
unperturbed horizon locally as a plane, and then the height $y$ of the
perturbed horizon above this plane as a function of distance $x$ along the
plane from the center of the fluctuation is the solution of the equation, 
$$
    {y\over\lam} 
    \left(1 
          + \sqrt {\left({x\over\lam}\right)^2 
          + \left({y\over\lam}\right)^2 }\;\right)
    \approx \left({\lam_0 \over \lam }\right)^3 .
$$
Thus, like the height, the lateral profile also depends only on the ratio
$R/\lam^3$.  When plotted, this profile looks like a smooth bump which, for
$\lam\sim\lam_0$, is about as wide as it is high.  (For all values of
$\lam/\lam_0$ the width of the bulge is comparable to the greater of its
height $h$ and the fluctuation radius $\lam$.)

\subsection { A self-similar wrinkling for $\lam\lto\lam_0$? }

To summarize, the size and shape of the horizon distortion induced by our
field fluctuation depends on the ratio $\lambda / \lambda_0$.  For
$\lambda\gg\lambda_0$ the fluctuation raises a bulge much smaller than
itself, whereas for $\lambda\ll\lambda_0$ the bulge is (nominally) much larger.
In particular, the deformation of the shape of the horizon, becomes
comparable in size to the fluctuation itself precisely when
$\lambda\sim\lambda_0$.  
It turns out that these conclusions do not depend on the specific profile
chosen for the effective mass density attributed to the fluctuation.  A
point mass leads to the same picture, as does a dipolar source with
vanishing total energy (perhaps a more appropriate model of a virtual
fluctuation of a quantum field).

On the other hand, the total neglect of the finite lifetime of the
fluctuation, and in particular of the attendant retardation effects, seems
a more serious matter.  We can assume (again on dimensional grounds) that
the fluctuation has a lifetime of order $\lam$, but it is not so obvious
how to take this into account in our Newtonian approximation.  One approach
is simply to imagine that the gravitational force due to the field
fluctuation is present only during its lifetime; or one could imagine in
addition that the force, while it exists, extends only a distance $\lam$
from the fluctuation.  With the first modification, the weakly coupled
fluctuations ($\lam\gto\lam_0$) behave basically as before, but for
$\lam\lto\lam_0$ the horizon distortions now remain at a height $\lam_0$
rather than growing indefinitely big; however, even this height far exceeds
the fluctuation size when $\lam\ll\lam_0$.  With the second modification
added in, it is plain that the bulge size can never exceed $\lam$ itself,
consistent with the intuition that the influence of a fluctuation should
not extend much beyond its immediate vicinity when retardation effects are
incorporated properly.

Thus, it seems plausible that the deformations in the horizon due to field
fluctuations of size $\lambda$ are actually of size $\lambda$ themselves,
for all $\lambda \lto \lambda_0 \sim (M l_p^2)^{1/3}$.  The resulting
horizon geometry could be described as ``fractal'' (meaning self-similar)
on scales between $l_p$ and $(M{l_p}^2)^{1/3}$ (it being doubtful whether
spacetime exists as a continuous manifold at all, on scales below $l_p$).
Perhaps one could also interpret this wrinkling of the horizon as a quantum
blurring of its location which effectively thickens it from a 2-dimensional
surface into a shell of thickness $\lam_0$.  In principle there is no limit
to how large this wrinkling could grow if sufficiently massive black holes
were available, but the prospect of human-sized distortions in the shape
fades when one plugs in the numbers: on a solar mass black hole, for
example, the bumps would only reach a scale of around $10^{-20}$~cm, and
for them to attain a size of even 1~cm, a black hole of the unheard of mass
of $10^{91}$~grams would be called for.

\section{ Implications and questions }

\subsection{ Implications }

 We can now tender a tentative answer to our question of how strongly the
horizon couples to the fluctuations of quantum fields (presumably including
the graviton field) propagating in its neighborhood.  To the extent that
the preceding analysis is a good guide, the answer is that the coupling is
weak on scales $\lam\gg\lam_0$ but strong in the opposite case.  The
implication of this for entanglement entropy is that the approximation of
quantum fields propagating in a fixed background geometry is unsuitable for
$\lam\lto\lam_0$, which means in turn that we are at present unable to
estimate reliably the magnitude of the entanglement entropy (or even to
define it) in this regime.  But if we limit ourselves to modes for which
$\lam\gto\lam_0$, we obtain only an entropy of magnitude
$$
      S_{entangle} 
           \sim { A\over(\lam_0)^2} 
           \sim {A/l_p^2\over (R/l_p)^{2/3}}
           \ll {A / l_p^2} .
$$
Hence entanglement entropy (at least the portion of it that we understand)
cannot provide more than a small fraction of the total horizon entropy.  

If the full entanglement entropy were indeed small, that would resolve the
species-dependence problem (to the extent that any problem remains), but it
would also force us to seek a different source for the bulk of the horizon
entropy.  Of course the horizon fluctuations we have just derived are
themselves such a source [6], and they should provide
approximately the right amount of entropy as well (assuming, as always, a
cutoff at around the Planck length), because they are equally as numerous
as the field fluctuations to which they correspond, and which in some sense
they replace.

\subsection{  Questions concerning a fully relativistic treatment }

To what extent could our ``improved Newtonian'' computation be repeated in
the context of full general relativity, and to what extent would we expect
to arrive at the same conclusions if we did repeat it?  Indeed, what
exactly do we mean here by ``the same conclusions''?  I am not going to try
to answer these questions now, but only to amplify them somewhat in the
following list.

\noindent $\bullet$ How should we model the field fluctuations?

In the Lorentzian context, would a (smeared out) {\it energy loop} be a
suitable model of the effective stress-energy tensor $T^{ab}$ of a field
fluctuation, since it would be conserved?  (The Newtonian equivalent could
be an extended mass dipole.)  But wouldn't we really need a Lorentz
invariant distribution of such loops?

Or, rather than trying to {\it model} fluctuations in $T^{ab}$, could we
just use the (renormalized) operator $\hat{T^{ab}}$ itself, and compute the
induced horizon distortions directly from it.  Perhaps this could be
accomplished via the Raychaudhuri equation.

\noindent $\bullet$  Can we compute the horizon distortion in a graviton
picture? 

Here the idea would be to translate the quantized linear fluctuations in
the metric (gravitons) directly into horizon distortions, and analyze the
latter using the correlation functions of the graviton field.  This would
be complementary to the kind of computation performed above, because
effective stress-energies wouldn't be involved at all.

\noindent $\bullet$ How to handle the non-linear regime?

Only a question without any indication of an answer for now --- but a
crucial question since it is just those fluctuations (with
$\lambda\sim\l_p$) that contribute most to the entropy for which a linear
approximation is least likely to be adequate.

\noindent $\bullet$ Can we find the horizon shape ``thermodynamically''?

The idea here would be to {\it assume} that the fluctuation formula
$probability\propto e^{\Delta{S}}$ (valid at fixed energy) works as usual
for black holes, and use it to define a probability distribution on the
space of all initial data for the classical Einstein equations with a fixed
energy, but varying horizon area.  The only horizon distortions with
non-negligible probability would then be those with ${\Delta{S}}\sim{-1}$,
or equivalently $\Delta{A}\sim{-l_p^2}$.  One could then ask whether the
fractal shapes suggested above would emerge as the most probable
configurations in this non-dynamical, ``thermodynamic'' approach.

\noindent $\bullet$ Can we define a horizon dynamics?

If one could isolate an approximately autonomous set of dynamical equations
governing the time-development of the horizon in classical general
relativity, then one could try to ``quantize'' these equations, and thence
to find --- and compute the entropy of --- a suitable quantum state
representing the horizon in ``internal equilibrium''.  Such a state would
presumably be mixed because the dynamics (presumably) would be dissipative
(like that of the damped sound modes in our gas example).

Another approach might be to interpret the quasi-normal black hole modes as
horizon oscillators (sensible?), and then attempt to compute their entropy
from their damping constants.  Unlike the first suggestion, this one
obviously would be restricted to linearized fluctuations about stationary
black holes.

\noindent $\bullet$ Is the Newtonian picture frame-dependent or modified by
the gravitational redshift? 

Even if we accept the conclusions of our Newtonian calculation, there is
the question of how to interpret the ``thickening'' of the horizon by
$\Delta{r}\sim\lam_0$.  Does $\Delta{r}$ translate into a Schwarzschild
coordinate distance or a proper distance or something else (and if a proper
distance then does the reference frame matter, given that the horizon is a
null surface)?  Also, does the general relativistic red shift modify our
estimate of $\lam_0$?  There is some indication from both these sides, that
$\lam_0\sim{M}^{1/3}$ might change to $\lam_0\sim{M}^{1/2}$, in a generally
relativistic treatment.

Well, if we can answer some of these questions, then we should gain a much
better conception of the small-scale structure of the horizon; and that in
turn should allow us to make a more definite assertion than we can at
present, about whether the finiteness of a black hole's entropy necessarily
entails a fundamental spacetime discreteness.


\bigskip\noindent

In conclusion, I would especially like to thank the other participants at
the conference for their stimulating questions and suggestions, during and
after my talk.

This research was partly supported by NSF grant PHY-9600620.

\ReferencesBegin

[1] 
Sean A.~Hayward,
``General laws of black-hole dynamics'',
  {\it Phys. Rev. D} {\bf 49}:6467-6474 (1994)

[2] 
J.D.~Bekenstein,                           
 ``Do we understand black hole entropy?'',
   To appear in the Proceedings of the 7th Marcel Grossmann Meeting (MG7), 
   held  Stanford, July 24--30, 1994
   $<$e-print archive: gr-qc/9409015$>$

[3]  
 R.D.~Sorkin, 
``Toward an Explanation of Entropy Increase in the
    Presence of Quantum Black Holes'',
  {\it Phys. Rev. Lett.} {\bf 56}, 1885-1888 (1986)

[4] 
  R.D.~Sorkin,
``The Statistical Mechanics of Black Hole Thermodynamics'',
  in R.M. Wald (ed.) {\it Black Holes and Relativistic Stars}, 
  (U. of Chicago Press, to appear)

[5] 
L.~Bombelli, J.~Lee, D.~Meyer and R.D.~Sorkin, 
``Spacetime as a Causal Set'', 
  {\it Phys. Rev. Lett.} {\bf 59}:521-524 (1987)

[6] 
 L.~Bombelli,  R.K.~Koul, Lee~J. and R.D.~Sorkin, 
``A Quantum Source of Entropy for Black Holes'', 
  {\it Phys. Rev.} {\bf D34}, 373-383 (1986);
  see also [8] below.

[7] 
Rafael D.~Sorkin and Sumati Surya,
``Why do instantons always yield area as entropy?'',                 
 (in preparation);
see also
C.~Carlip and C.~Teitelboim,
``The off-shell black hole'',
 {\it Class. Quant. Grav.} {\bf 12}: 1699 (1995)

[8] 
 R.D.~Sorkin, 
``On the Entropy of the Vacuum Outside a Horizon'',
  in B. Bertotti, F. de Felice and A. Pascolini (eds.),
  {\it Tenth International Conference on General Relativity and Gravitation
  (held Padova, 4-9 July, 1983), Contributed Papers}, 
  vol. II, pp. 734-736
  (Roma, Consiglio Nazionale Delle Ricerche, 1983)

[9] 
%
%
M.~Srednicki, 			
``Entropy and Area'',
 {\it Phys.~Rev.~Lett.} {\bf 71}:666-669 (1993)
 $<$ e-Print Archive: hep-th/9303048 $>$;
V.~Frolov,
``Black Hole Entropy and Physics at Planckian Scales'',
 University of Alberta Preprint (1995)
 $<$ e-Print Archive: hep-th/9510156 $>$

[10] 
L.~Susskind and J.~Uglum, 
``Black hole entropy in canonical quantum gravity and superstring theory''
  {\it Phys. Rev.} D {\bf 50}:2700 (1994)
 $<$ e-Print Archive: hep-th/9401070 $>$

[11] 
 R.D.~Sorkin,
``Two Topics concerning Black Holes: 
   Extremality of the Energy, Fractality of the Horizon'',
   in S.A.~Fulling (ed.), 
   {\it Proceedings of the Conference on Heat Kernel Techniques and Quantum
    Gravity, held Winnipeg, Canada, August, 1994}, pp. 387-407
   (Discourses in Mathematics and its Applications, \#4,) 
   (University of Texas Press, 1995)
   (electronic archive designator: gr-qc/9508002)

\end